\documentclass[pre,showpacs,preprintnumbers,twocolumn,amsmath,amssymb]{revtex4}

\usepackage{graphicx}
\usepackage{epsfig}

\newcommand{\MB}{\ensuremath{{\rm MB}}}
\newcommand{\eq}{\ensuremath{{\rm eq}}}
\newcommand{\err}{\ensuremath{\prime}}
\newcommand{\neqq}{\ensuremath{{\rm neq}}}
\newcommand{\errneq}{\ensuremath{\prime\prime}}
\newcommand{\Ec}{\ensuremath{{\rm Ec}}}
\newcommand{\Nu}{\ensuremath{{\rm Nu}}}
\newcommand{\Ra}{\ensuremath{{\rm Ra}}}

\begin{document}
\title{Lattice Boltzmann method for thermal flow simulation on standard lattices}

\author{Nikolaos I. Prasianakis}\email{prasianakis@lav.mavt.ethz.ch}
\affiliation {Aerothermochemistry and Combustion Systems Lab,
Institute of Energy Technology, ETH Zurich, 8092 Zurich,
Switzerland}

\author{ Iliya V. Karlin\footnote{Corresponding author}}\email{karlin@lav.mavt.ethz.ch}
\affiliation {Aerothermochemistry and Combustion Systems Lab,
Institute of Energy Technology, ETH Zurich, 8092 Zurich,
Switzerland}

\begin{abstract}

The recently introduced consistent lattice Boltzmann model with
energy conservation [S. Ansumali, I.V. Karlin, Phys. Rev. Lett.
{\bf 95}, 260605 (2005)] is extended to the simulation of thermal
flows on standard lattices. The two-dimensional thermal model on
the standard square lattice with nine velocities is developed and
validated in the thermal Couette and Rayleigh-B\'{e}nard natural
convection problems.

\end{abstract}
\maketitle

\section{Introduction}

The lattice Boltzmann (LB) method is a powerful approach to
hydrodynamics, with applications ranging from large Reynolds
number flows to flows at a micron scale, porous media and
multi-phase flows \cite{SucciBook}. The LB method solves a fully
discrete kinetic equation for populations $f_i(x,t)$, designed in
a way that it reproduces the target equations of continuum
mechanics in the hydrodynamic limit. Populations correspond to
discrete velocities $c_i$, $i=1,\dots, N$, which fit into a
regular spatial lattice with the nodes $x$. This enables a simple
and  efficient `stream-along-links-and-equilibrate-at-nodes'
realization of the LB algorithm.

In the case of incompressible flows, where the target equations
are Navier-Stokes equations at low Mach number, the LB method
proves to be competitive to conventional methods of computational
fluid dynamics \cite{SCI}. In that case, the LB models on the
so-called standard lattices with a relatively small number of
velocities ($N=9$ in two dimensions, see Fig.\ \ref{Fig1},  and
$N=15,19,27$ in three) are available and most commonly used. In
this paper, we will follow the usual nomenclature and indicate the
models as $DMQN$ where $M=2,3$ is the spatial dimension and $N$
the number of the discrete velocities of the model.

However, situation is quite different with LB models for
compressible thermal flows. In spite of a  number of recent
suggestions, a commonly accepted LB model for thermal flows has
not yet been established, to the best of our knowledge. A reason
for such a difference between the isothermal and thermal cases is
mainly because it is not straightforward to incorporate the
temperature into
 the lattice equilibrium when using the
 standard lattices, and simultaneously to satisfy a number of conditions for recovering the
 Navier-Stokes and Fourier equations for compressible flows \cite{RMP}.
 At present, two major approaches to constructing thermal LB models can be
 distinguished. In the first approach, lattices with a larger
 number of discrete velocities or off-lattice velocity sets are
 considered to enable local energy conservation and isotropy
 \cite{Ansumali03,Chikatamarla06}. In the second approach, two
 copies of the standard lattices are considered, one of which
 is designed to treat the density and momentum and the other - the
 energy density. The standard isothermal LB model on the first
 lattice is considered, and the coupling between the lattices is
 enhanced by introducing force terms in order to recover viscous
 heat dissipation \cite{HeChenDoolen98}. However, both these approaches
 inevitably deal with more discrete speeds than the standard single-lattice
 LB models which leads to reducing efficiency.

Recently, a so-called consistent LB model has been introduced in
Ref.\  \cite{Ansumali05}. It has been shown that it is possible to
construct the LB model with energy conservation on the standard
lattices in such a way that the spurious bulk viscosity of
isothermal LB models is eliminated, and that the Navier-Stokes and
Fourier equations are recovered once the temperature is varied in
a small neighborhood of a reference temperature. However, the
consistent LB model is limited to weakly compressible flows, and
is not yet sufficient to simulate thermal flows. The reason for
that is the low symmetry of the standard lattices which leads to
accumulation of deviations of the resulting hydrodynamic equations
from their Navier-Stokes and Fourier counterparts. It should be
noted, however, that the consistent LB model with energy
conservation on standard lattices should be considered as a better
and more natural starting point for a development of the thermal
LB model as compared to the isothermal model.

In this paper, we complete the program of constructing the thermal
LB model on standard lattices as an extension of the consistent LB
model \cite{Ansumali05}. A new thermal lattice Boltzmann model is
derived on the most commonly used $D2Q9$ lattice as follows: In
section \ref{SecLE}, we revisit the derivation of the local
equilibrium of the thermal model, using the method of guided
equilibrium \cite{Karlin98}. This enables to enhance Galilean
invariance of the consistent LB method for large temperature
variations. We shall also identify the remaining deviations in the
higher-order moments due to the lattice constraints. In section
\ref{SecDev}, the impact of these deviations on the hydrodynamic
equations is identified (algebraic details of the derivation are
presented in the Appendix. In section \ref{SecCorr}, additional
terms are introduced in the Boltzmann equation in such a way that
all the deviations of the hydrodynamic equations from their
Navier-Stokes and Fourier counterparts are eliminated. In
addition, further terms are introduced to represent external
forces, as well as to tune the Prandtl number of the model. In
section \ref{SecDiscr}, the discretization in time and space of
the kinetic equation of section \ref{SecCorr} is performed, and a
simple numerical algorithm is outlined. In section \ref{SecNum},
numerical examples are provided. Simulations of the thermal
Couette flow and of the Rayleigh-B\'{e}nard natural convection
show excellent agreement with theoretical results. The numerical
implementation is summarized in a compact form in section
\ref{SecSum}. Finally, results are discussed in section
\ref{SecConcl}.

\section{Local equilibrium}\label{SecLE}
\subsection{Consistent lattice Boltzmann method}
\begin{figure}[tbp]
    \centering
        \includegraphics[width=0.22\textwidth]{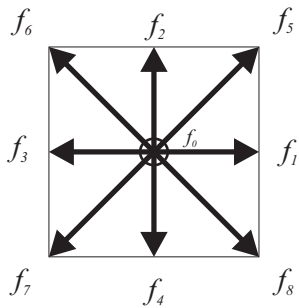}
    \caption{The $D2Q9$ velocity set. }
    \label{Fig1}
\end{figure}

We begin with reminding the construction of the local equilibrium
in the consistent lattice Boltzmann method
\cite{Ansumali05,Prasianakis06} on the planar square lattice with
nine velocities $c_{i\alpha}$, $i=0,\dots, 8$ (see Fig.\
\ref{Fig1}):
\begin{align}
\label{velset}
\begin{split}
c_x &= \left\{0, 1, 0, -1, 0, 1,-1,-1,1\right\}\\
c_y &= \left\{0, 0, 1,  0, -1, 1,1,-1,-1\right\}
\end{split}
\end{align}
Equilibrium populations are obtained from minimization of the
entropy function $H$ under constraints provided by
 local conservation laws.
 For the $D2Q9$ model, the entropy function has the form
 \cite{DHT},
\begin{equation}
H = \sum_{i=0}^{8} f_i \ln \left(\frac{f_i}{W_i}\right),
\label{HfuncBoltz}
\end{equation}
where the weights $W_i$ are:
\begin{equation}
W =\frac{1}{36} \left\{16, 4, 4,  4, 4, 1,1,1,1\right\}.
\end{equation}
In the consistent lattice Boltzmann method \cite{Ansumali05}, the
local conservations include mass, momentum and energy,
\begin{align}
\begin{split} \sum\limits_{i = 0}^8 {f^{\eq}_i }  &= \rho, \\
\sum\limits_{i = 0}^8 {c_{i\alpha } f_i^{\eq} } & =
j_\alpha,\\
\sum\limits_{i = 0}^8 {c_i^2 f_i^{\eq} }  &= 2\rho T +
\frac{j^2}{\rho} \label{Mconstr}
\end{split}
\end{align}
where $\rho$, $j_{\alpha}=\rho u_{\alpha}$ and $T$ are the
density, momentum and temperature fields, correspondingly. Note
that the consistent lattice Boltzmann construction includes energy
conservation among the constraints (\ref{Mconstr}). Note that this
is at variance with the standard isothermal lattice Boltzmann
method on the same lattice \cite{LBGK2} where the entropy function
is minimized under the constraints of mass and momentum only
\cite{DHT}.

Derivation of the equilibrium populations as a minimizer of the
entropy function (\ref{HfuncBoltz}) under constraints
(\ref{Mconstr}) can be found in Refs.\
\cite{Ansumali05,Prasianakis06}, and is not reproduced here. We
remind that at the reference temperature $T_0=1/3$,  the
consistent lattice Boltzmann model recovers the
Navier-Stokes-Fourier hydrodynamic equations and results in the
nonequilibrium pressure tensor without a spurious bulk viscosity
of the standard lattice Boltzmann model. However, the consistent
lattice Boltzmann model itself is not yet capable of simulating
accurately thermal flows with significant temperature and density
variations.
 In particular,
the equilibrium pressure tensor of the consistent lattice
Boltzmann model shows deviations of the order of  $u^2 \Delta T$
(where $\Delta T$ is a deviation from the reference temperature)
from the target Maxwell-Boltzmann form,
\begin{equation}
P_{\alpha \beta}^{\MB} =\rho T \delta _{\alpha \beta} +
\frac{j_{\alpha} j_{\beta}}{\rho}. \label{MBpressure}
\end{equation}
 Hence, as the first step towards establishing the lattice
Boltzmann method for thermal flow simulations, we need to modify
the construction of the equilibrium which would remove this
deviation from the pressure tensor.

\subsection{Guided local equilibrium}

In order to remove the aforementioned deviation in the equilibrium
pressure tensor, we use the method of guided equilibrium
introduced in \cite{Karlin98} for a generic lattice Boltzmann
model. Following \cite{Karlin98}, we minimize the entropy function
(\ref{HfuncBoltz}) under an extended set of conditions which
includes the local conservations (\ref{Mconstr}) and the condition
which stipulates that the equilibrium pressure
 tensor $P_{\alpha \beta}^{\eq}$ is in the MB form (\ref{MBpressure}),
\begin{equation}
\sum\limits_{i = 0}^8 {c_{i\alpha } c_{i\beta } f_i^{\eq} }  =
\rho T\delta _{\alpha \beta }  + \frac{j_\alpha  j_\beta}{\rho}.
\label{GuidedP}
\end{equation}
Minimization of the entropy function (\ref{HfuncBoltz}) under the
extended list of constraints, (\ref{Mconstr}) and (\ref{GuidedP})
leads to the  equilibrium populations of the form
\begin{equation}
f_i^{\eq}  = w_i \exp \{ \mu  + \zeta _\alpha c_{i\alpha} + \chi
_{\alpha\beta} c_{i\alpha} c_{i\beta} + \gamma c_i^2 \},
\label{fieq}
\end{equation}
where summation convention in two repeated spatial indices is
assumed, and where $\mu$, $\zeta _\alpha$, $\chi _{a \beta}$,
$\gamma$ are Lagrange multipliers of the constraints. Their values
are found upon substitution of (\ref{fieq}) into (\ref{Mconstr})
and (\ref{GuidedP}). Equilibrium at zero velocity ($j_{\alpha}=0$)
can be found exactly and coincides with the one reported in
\cite{Ansumali05,Prasianakis06}. Equilibria at non-zero velocity
are then derived by perturbation around the zero-velocity state,
and $f_i^{\eq}$ are represented in terms of a series in the
momentum. For what will follow, expansion up to the fourth order
in velocity is required. This solution procedure is quite
standard, hence, we reproduce only the result in terms of the
velocity $u_\alpha=j_\alpha/\rho$:

\begin{align}
\begin{split}
\label{eqpop}
f_0^{\eq}  &= \rho (T + u_x^2 - 1)(T + u_y^2 - 1),     \\
f_1^{\eq}  &= \frac{\rho}{2}(T + u_x  + u_x^2 )(1 - T - u_y^2 ),  \\
f_2^{\eq}  &= \frac{\rho}{2}(1 - T - u_x^2 )(T + u_y  + u_y^2 ), \\
f_3^{\eq}  &= \frac{\rho}{2}(T - u_x  + u_x^2 )(1 - T - u_y^2 ), \\
f_4^{\eq}  &= \frac{\rho}{2}( 1 - T - u_x^2 )(T - u_y  + u_y^2 ), \\
f_5^{\eq}  &=\frac{\rho}{4}(T + u_x  + u_x^2 )(T + u_y  + u_y^2 ),\\
f_6^{\eq}  &=\frac{\rho}{4}(T - u_x  + u_x^2 )(T + u_y  + u_y^2 ),\\
f_7^{\eq}  &=\frac{\rho}{4}(T - u_x  + u_x^2 )(T - u_y  + u_y^2),\\
f_8^{\eq}  &=\frac{\rho}{4}(T + u_x  + u_x^2 )(T - u_y  + u_y^2 ).
\end{split}
\end{align}

Note that equilibrium populations at zero velocity remain positive
for temperature values $0<T<1$. The equilibrium pressure tensor
satisfies the MB relation per construction,
\begin{equation}
P_{\alpha\beta}^{\eq}=P_{\alpha\beta}^{\MB}.
\end{equation}
The properties of the higher-order  moments in the equilibrium
(\ref{eqpop}) are studied in the next section.

\subsection{Deviations in higher-order moments}

Apart from the equilibrium pressure tensor, also the third- and
fourth-order moments of the equilibrium must satisfy the
Maxwell-Boltzmann relations in order to recover the Navier-Stokes
and the temperature equation in the hydrodynamic limit,
\begin{align}
\begin{split}
Q_{\alpha \beta \gamma }^{\MB}  &= T(j_\alpha  \delta _{\beta
\gamma }  + j_\beta  \delta _{\alpha \gamma }  + j_\gamma  \delta
_{\alpha \beta } )
+ \frac{{j_\alpha j_\beta  j_\gamma  }}{{\rho ^2 }},  \\
R_{\alpha \beta }^{\MB}  &= T\left( {\frac{{j^2 }}{\rho } + 4T}
\right)\delta _{\alpha \beta }  + 6T\frac{{j_\alpha  j_\beta
}}{\rho } + \frac{{j_\alpha j_\beta  j^2 }}{{\rho ^3 }}.
\end{split}
\end{align}
Using the equilibrium populations (\ref{eqpop}) to evaluate the
functions
\begin{align}
\begin{split}
Q_{\alpha \beta \gamma }^{\eq}  &= \sum_{i = 0}^8
{c_{i\alpha } c_{i\beta } c_{i\gamma } f_i^{\eq} }, \\
R_{\alpha \beta }^{\eq}  &= \sum_{i = 0}^8 {c_{i\alpha } c_{i\beta
} c_i^2 f_i^{\eq} },
\end{split}
\end{align}
and introducing, for a generic moment $M$, a deviation $M^{\err}$
of its equilibrium value $M^{\eq}$ from the Maxwell-Boltzmann
value $M^{\MB}$,
\begin{equation}
M^{\eq}  = M^{\MB}  + M^{\err}, \label{notat1}
\end{equation}
we find the following deviations of the third- and fourth-order
moments:
\begin{align}
\begin{split}
\label{Error}
Q_{xxx}^{\err}  &= j_x (1 - 3T) - \frac{j_x^3}{\rho^2},  \\
Q_{yyy}^{\err}  &= j_y (1 - 3T) - \frac{j_y^3}{\rho^2},  \\
R_{xx}^{\err} & = (1 - 3T)\rho T + (1 - 6T)\frac{{j_x^2 }}{\rho
}
- \frac{{j_x^4 }}{{\rho ^3 }},\\
R_{yy}^{\err} & = (1 - 3T)\rho T + (1 - 6T)\frac{{j_y^2 }}{\rho
}
- \frac{{j_y^4 }}{{\rho ^3
}},\\
R_{xy}^{\err} & = 2(1 - 3T)\frac{{j_x j_y }}{\rho } - \frac{{j_x
j_y j^2 }}{{\rho ^3 }}.
\end{split}
\end{align}
Note that the off-diagonal elements of the third-order moment
$Q_{\alpha\beta\gamma}^{\eq}$ already satisfy the Maxwell-Boltzmann
relations,
\begin{equation}
Q_{xxy}^{\eq}=Q_{xxy}^{\MB},\ Q_{yyx}^{\eq}=Q_{yyx}^{\MB}.
\end{equation}
Thus, the deviation of the contracted third-order moment,
$q_{\alpha}^{\eq}=\sum_{i=0}^{8}c_{i\alpha}c_i^2 f_i^{\eq}$ (the
energy flux) is due to the deviation of the diagonal elements,
\begin{equation}
q_{\alpha}^{\err}=Q_{\alpha\alpha\alpha}^{\err}. \label{Errq}
\end{equation}

Deviations of the diagonal components of the third-order moments
are well known for the low-symmetry $D2Q9$ lattice, and stem from
the fact that the velocity set satisfies a relation,
$c_{i\alpha}^3=c_{i\alpha}$. This lattice constraint precludes
construction of equilibrium different from (\ref{eqpop}) in such a
way that also the energy flux would be guided by the
Maxwell-Boltzmann relation.

In the next section, we shall identify the impact of the
deviations (\ref{Error}) on the hydrodynamic equations. Once this
will be done, we shall remove the anomalous terms in the
hydrodynamic equations by introducing correction terms into the
kinetic equation.

\section{Deviations in the hydrodynamic equations}\label{SecDev}
\subsection{Bear BGK model}

 In this section, the impact of deviations of the moments (\ref{Error})
 on the hydrodynamic equations will be identified
 using the single relaxation time Bhatnagar-Gross-Krook (BGK)
 model for the collision integral:
\begin{eqnarray}
\partial _t f_i  + c_{i\alpha} \partial _{\alpha} f_i  =
- \frac{1}{\tau }(f_i  - f_i^{\eq} ), \label{KE1}
\end{eqnarray}
where $\tau$ is the relaxation time, and the equilibrium
populations are given by Eq.\ (\ref{eqpop}). The hydrodynamic
limit of the BGK kinetic equation (\ref{KE1}) can be studied via
the Chapman-Enskog expansion. Details of the Chapman-Enskog
analysis are given in the Appendix \ref{AppendixA}. Here we shall
summarize the results of this analysis.
\subsection{Deviations in the momentum equation}
To the first order of the Chapman-Enskog expansion,  the momentum
equation reads:
\begin{equation}
\partial _t j_\alpha  + \partial _\beta  P_{\alpha \beta }^{\MB}
+\partial _\beta  P_{\alpha \beta }^{\neqq} = - \partial_\gamma
{{P}_{\alpha \gamma }^{\errneq} }, \label{MomeqTELBM}
\end{equation}
where $P_{\alpha \beta }^{\neqq}$ is the nonequilibrium pressure
tensor in the form required for the Navier-Stokes equation,
\begin{equation}
P_{\alpha \beta }^{\neqq}=-\tau\rho T\left[\partial _\alpha
\left(\frac{{j_\beta }}{\rho }\right) + \partial _\beta
\left(\frac{{j_\alpha }}{\rho }\right) - \delta _{\alpha \beta
}\partial _\gamma \left(\frac{{j_\gamma  }}{\rho } \right)
\right],
\end{equation}
while $P_{\alpha \beta }^{\errneq}$ is the deviation of the
nonequilibrium pressure tensor from $P_{\alpha \beta }^{\neqq}$:
\begin{widetext}
\begin{align}
\begin{split}
\label{pygamma}
\partial_\gamma {{P}_{x\gamma }^{\errneq}}&=-\frac{\tau}{2}\partial_x \left[{\partial_x \left({j_x(1-3T)
-\frac{{j_x^3}}{\rho}} \right) - \partial_y
\left({j_y(1-3T)-\frac{{j_y^3}}{\rho}}\right)}\right], \\
\partial _\gamma {{P}_{y\gamma }^{\errneq}}&=-\frac{\tau}{2}\partial_y \left[ {\partial_y \left( {j_y (1 - 3T) -
\frac{{j_y^3 }}{\rho }} \right) - \partial_x \left( {j_x (1 - 3T)
- \frac{{j_x^3 }}{\rho }} \right) } \right].
\end{split}
\end{align}
\end{widetext}
In order to recover the Navier-Stokes equation, we will need to
introduce additional terms into Eq.\ (\ref{KE1}) in order to
annihilate the right hand side of the momentum equation
(\ref{MomeqTELBM}) (see section \ref{SecCorr}).

\subsection{Deviations in the energy equation}
Similarly, the energy density equation reads,
\begin{equation}
\partial _t \left( {\frac{{j^2 }}{\rho } + 2\rho T} \right)
 + \partial _\alpha  q_\alpha^{\MB}  +\partial _\alpha  q_\alpha ^{\neqq}=
  - \partial _\alpha  q_\alpha^{\err}  - \partial _\alpha  {q}_\alpha^{\errneq},  \label{nrgeq}
\end{equation}
where
\begin{equation}
q_\alpha ^{\neqq}  = -4\tau\rho T\partial _\alpha  T +
2\left(\frac{{j_\beta }}{\rho }\right)P_{\alpha \beta }^{\neqq},
\label{qaneqq}
\end{equation}
is the nonequilibrium energy flux required in the Fourier equation
for the energy density. The first term in (\ref{qaneqq}) is the
Fourier law of energy dissipation, whereas the second term
represents viscous dissipation.
 Again, terms in the right hand side of
(\ref{nrgeq}) express the deviation of the energy equation from
the required form. The first of these terms is given by Eqs.
(\ref{Errq}) and (\ref{Error}), while the result for the second
term, ${q}_\alpha^{\errneq}$, is given by the following
expression:

\begin{align}
\label{qaerr}
\begin{split}
q_{\alpha}^{\errneq} &=  3\tau \rho T\partial _\alpha  T  \\
& - \tau \left[ {\frac{3j_\alpha j_\alpha }{{\rho ^2 }}\partial
_\alpha(\rho T) + 3j_\alpha  T\partial _\beta \left(\frac{{j_\beta
}}{\rho }\right)
- \frac{3j_\alpha  j_\beta }{2\rho }\partial _\beta  T} \right] \\
& - \tau \left[ {(1 - 3T)\left( {\partial _\beta
\left(\frac{{j_\alpha j_\beta  }}{\rho }\right) - \frac{{j_\alpha
}}{{2\rho }}\partial _\beta
j_\beta } \right)} \right]\\
& - \tau \left[ {2\frac{j_\alpha ^3 }{\rho ^3 }\partial _\beta
j_\beta + \frac{3j_\alpha j_\alpha}{\rho ^2}\partial _\beta
\left(\frac{{j_\alpha j_\beta  }}{\rho }\right) + \frac{{j_\alpha
}}{{2\rho
}}\partial _\beta  \left(\frac{j_\beta^3 }{\rho ^2 }\right)} \right]\\
&+\tau \partial _\beta  e_{\alpha \beta } ^{\errneq},
\end{split}
\end{align}
where
\begin{align}
\label{EnoneqShort}
\begin{split}
e_{xx}^{\errneq} & =   \frac{j_x j_x}{\rho}  + \frac{j_x^4}{\rho^3},  \\
e_{yy}^{\errneq}  &=   \frac{j_y j_y}{\rho } + \frac{j_y^4}{\rho ^3},  \\
e_{xy}^{\errneq} & =   \frac{j_x j_y j^2}{\rho ^3}.  \\
\end{split}
\end{align}
It should be noted that, in certain situations, many terms in
(\ref{qaerr}) and (\ref{EnoneqShort}) can be safely neglected. For
example, deviations of the order $j^4$ affect the viscous heat
dissipation and can be ignored in cases where viscous heating is
unimportant. However, we shall use the full expressions
(\ref{qaerr}) and (\ref{EnoneqShort}) for the deviation in the
numerical implementation below.

The deviations in the hydrodynamic equations identified in this
section are due to the lattice constraints of the $D2Q9$ lattice,
and they cannot be removed by introducing a set of the equilibrium
distribution functions different from (\ref{eqpop}). These
deviations can be removed only by introducing correction terms
into the kinetic equation which we address in the next section.

\section{Correction of BGK equation}\label{SecCorr}
To this end, we have computed the deviations due to the lattice
constraints. In this section, an efficient way to remove these
deviations by  adding correction terms in the Boltzmann equation
is introduced:
\begin{eqnarray}
\partial_t f_i+c_{i\alpha}\partial_\alpha f_i=
-\frac{1}{\tau}(f_i-f_i^{\eq})+\Psi_i + \Phi_i.  \label{keqf}
\end{eqnarray}
The purpose of the $\Psi_i$ terms is to correct the momentum
equation, while the purpose of the $\Phi_i$ terms is to correct
the energy equation.
\subsection{Momentum equation correction terms $\Psi_i$}

We require that the $\Psi_i$ term affects only the momentum
equation and delivers there a term which compensates $-
\partial_\gamma {{P}_{\alpha \gamma }^{\errneq} }$. Thus,
\begin{align}
\begin{split}
 \sum_{i=0}^{8} {\Psi_i}  &= 0,\\
 \sum_{i=0}^{8}{c_{i\alpha } \Psi_i } & = \partial_\gamma P_{\alpha\gamma }^{\errneq},\\
 \sum_{i=0}^{8}{c_i^2 \Psi_i }  &= 0,\\
 \sum_{i=0}^{8}{c_{i\alpha } c_{i\beta } \Psi_i }  &= 0,\\
 \sum_{i=0}^{8}{c_{i\alpha } c_i^2 \Psi_i }  &= 0, \\
 \sum_{i=0}^{8}{c_{i\alpha } c_{i\beta } c_i^2 \Psi_i }  &= 0.
\end{split}
\end{align}
Among these equations only nine are independent, and the unique
solution is:
\begin{align}
\begin{split}
\label{PsiResult}
\Psi_0 & =0,\\
\Psi_1 & = \partial_\gamma P_{x\gamma }^{\errneq},\\
\Psi_2 & =\partial_\gamma P_{y \gamma }^{\errneq},\\
\Psi_3 & =-\partial_\gamma P_{x \gamma }^{\errneq},\\
\Psi_4 & =-\partial_\gamma P_{y \gamma }^{\errneq},\\
\Psi_5 & =-\frac{1}{4}\left(\partial_\gamma P_{x\gamma}^{\errneq}+
                            \partial_\gamma P_{y \gamma }^{\errneq}\right),\\
\Psi_6 & =\frac{1}{4}\left(\partial_\gamma P_{x\gamma }^{\errneq}-
                        \partial_\gamma P_{y \gamma }^{\errneq}\right), \\
\Psi_7 & =\frac{1}{4}\left(\partial_\gamma P_{x\gamma }^{\errneq}+
                        \partial_\gamma P_{y \gamma }^{\errneq}\right),\\
\Psi_8 & =\frac{1}{4}\left(-\partial_\gamma P_{x\gamma
}^{\errneq}+
\partial_\gamma P_{y \gamma }^{\errneq}\right).
\end{split}
\end{align}

\subsection{Energy equation correction terms $\Phi_i$}

Similarly, it is required that $\Phi_i$ terms appear only in the
energy equation,
\begin{align}
\begin{split}
 \sum_{i=0}^{8}\Phi_i  &= 0,\\
 \sum_{i=0}^{8}c_{i\alpha} \Phi_i & = 0,\\
 \sum_{i=0}^{8}c_i^2 \Phi_i &= \partial _\alpha(q_\alpha^{\err}+q_\alpha ^{\errneq}), \\
 \sum_{i=0}^{8}c_{ix} c_{iy} \Phi_i & = 0,\\
 \sum_{i=0}^{8}c_{i\alpha}c_i^2 \Phi_i & = 0,\\
 \sum_{i=0}^{8}c_{i\alpha } c_{i\beta } c_i^2 \Phi_i & = 0.
\end{split}
\end{align}
The solution of that system is:
\begin{align}
\begin{split}
\label{PhiResult}
\Phi_0&= -\frac{3}{2}\partial _\alpha(q_\alpha^{\err}+q_\alpha^{\errneq}),\\
\Phi_1&=\frac{1}{2}{\partial _\alpha(q_\alpha^{\err}+q_\alpha^{\errneq})},\\
\Phi_2&=\frac{1}{2}\partial_\alpha(q_\alpha^{\err}+q_\alpha^{\errneq}),\\
\Phi_3&=\frac{1}{2}\partial_\alpha(q_\alpha^{\err}+q_\alpha^{\errneq}),\\
\Phi_4&=\frac{1}{2}\partial_\alpha(q_\alpha^{\err}+q_\alpha^{\errneq}),\\
\Phi_5&=-\frac{1}{8}\partial_\alpha(q_\alpha^{\err}+q_\alpha^{\errneq}),\\
\Phi_6&=-\frac{1}{8}\partial_\alpha(q_\alpha^{\err}+q_\alpha^{\errneq}),\\
\Phi_7&=-\frac{1}{8}\partial_\alpha(q_\alpha^{\err}+q_\alpha^{\errneq}),\\
\Phi_8&=-\frac{1}{8}\partial_\alpha(q_\alpha^{\err}+q_\alpha^{\errneq}).
\end{split}
\end{align}

\subsection{Variable Prandtl number}
 A thermal lattice Boltzman model should be able to simulate fluids with an  arbitrary Prandtl
 number. The Prandtl  number $\Pr$ is defined as the ratio of the viscosity and the thermal
 conductivity,
\begin{eqnarray}
\Pr=\frac{c_p \mu}{\kappa},
\end{eqnarray}
where $c_p$ is specific heat under a constant pressure. We here
suggest the following simple way to change the Prandtl number in
the present model by adjusting the correction term $\Phi_i$. In
order to do this, we first note that, with the use of the guided
equilibrium populations, the BGK model (\ref{KE1}) leads to
$\Pr=4$, same as the original consistent LB method
\cite{Prasianakis06}. Applying the correction for the energy
equation mentioned in the previous section, the Prandtl number
becomes $\Pr=1$, as in the standard continuous BGK model. This
happens because the correction compensates, among the others, the
term,
\[3\tau
\rho T
\partial_\alpha T,\]
 (first term in (\ref{qaerr})), responsible for the change of the thermal conductivity of
the model. Thus, arbitrary values of $\Pr$  can be obtained by
applying the following transformation in (\ref{qaerr}) and thereby
altering the counter-term $\Phi_i$ in is term:
\begin{eqnarray}
3 \tau
\rho T\partial _\alpha  T \to
\frac{{\left( {4 - \Pr } \right)}}{{\Pr }} \tau
\rho T\partial _\alpha  T. \label{prcorr}
\end{eqnarray}
Thus, with the simple transform (\ref{prcorr}) in the energy
correction term $\Phi_i$, kinetic equation  (\ref{keqf}) recovers
the hydrodynamic equations with a predetermined Prandtl number. In
section \ref{SecNum}, we shall validate this in the thermal
Couette flow for fluids with various Prandtl numbers.

\subsection{Gravity force}
Flows  subject to external forces, such as gravity, are a major
concern in many applications. In order to incorporate the force
$\rho g$, where $g$ is the acceleration due to gravity, it
suffices to apply the following transformation when calculating
the correction terms $\Psi_i$:
\begin{equation}
\partial_\beta P_{\alpha  \beta }^{\errneq} \to \partial_\beta P_{\alpha \beta }^{\errneq}
+ \rho g_\alpha. \label{gravcorr}
\end{equation}
Any additional physics can be incorporated into the present model
by similar considerations.
\subsection{Corrected hydrodynamic equations}
Taking into account all the aforementioned correction terms
$\Psi_i$ and $\Phi_i$ in the kinetic equation (\ref{keqf}), and
using again the Chapman-Enskog analysis, we obtain the
two-dimensional Navier-Stokes and Fourier hydrodynamic equations
for density, momentum and temperature:
\begin{widetext}
\begin{align}
\begin{split}
\label{NSF}
\partial _t \rho  &= - \partial_\gamma(\rho u_\gamma), \\
\partial _t u_\alpha  &= - u_\gamma  \partial_\gamma  u_\alpha -
\frac{1}{\rho }\partial_\alpha (\rho T) + \frac{1}{\rho
}\partial_\gamma \left[ \tau \rho T \left ( \partial _\alpha
 u_\gamma  + \partial _\gamma
 u_\alpha  - \partial _\kappa  u_\kappa  \delta _{\alpha \gamma
} \right) \right] + \rho g_\alpha,    \\
\partial _t T &=  - u_\alpha \partial _\alpha T - T\partial _\alpha u_\alpha
- \frac{1}{\rho }\left[ {\partial _\alpha \left( \frac{2}{\Pr} \tau \rho T\partial _\alpha T \right)
 + \left( \partial _\gamma  u_\alpha \right)   \tau \rho T \left ( \partial _\alpha u_\gamma  + \partial _\gamma
 u_\alpha  - \partial _\kappa  u_\kappa  \delta _{\alpha \gamma} \right) }
 \right].
 \end{split}
 \end{align}
\end{widetext}
The fluid corresponds to the ideal gas equation of state, $p=\rho
T$, with specific heats $c_v=1.0$ and $c_p=2.0$, in lattice units,
and with the adiabatic exponent $\gamma = c_p/c_v = 2.0$. The
viscosity coefficient $\mu$ and thermal conductivity $\kappa$ can
be identified as
\begin{eqnarray}
\mu= \tau \rho T,\ \kappa=\frac{2}{\Pr} \tau \rho
T.\label{transportcoefficients}
\end{eqnarray}
The Prandtl number is a tunable parameter and can take arbitrary
values. A wide range of gases and  fluids whose physics is
described by equations (\ref{NSF}) can be simulated. We now
proceed with a time-space discretization of the kinetic equation
(\ref{keqf}).

\section{Lattice Boltzmann discretization}\label{SecDiscr}
Derivation of   the lattice Boltzmann scheme for the kinetic
equation (\ref{keqf}) proceeds essentially along the lines of
Ref.\ \cite{HeChenDoolen98}. First, we integrate over the time
step $\delta t$ along the characteristics:
\begin{widetext}
\begin{equation}
f_i (x + c_i \delta t ,t + \delta t) = f_i(x,t) +
\int\limits_t^{t+\delta t}
{\frac{1}{\tau}(f_i^{\eq}(t')-f_i(t'))dt'} +
\int\limits_t^{t+\delta t} {\Psi_i(t')dt'}  +
\int\limits_t^{t+\delta t} {\Phi_i(t')dt'}. \label{disckeqf}
\end{equation}
\end{widetext}
The time integrals of the collision term as well  as of the
correction terms is evaluated  using the second-order accurate
trapezoidal rule. Second, in order to establish a semi-implicit
scheme, the following transformation is applied
\cite{HeChenDoolen98}:
\begin{eqnarray}
g_i &=& f_i  + \frac{{\delta t}}{2\tau}(f_i-f_i^{\eq}) -
\frac{{\delta t}}{2}[\Psi_i  + \Phi_i] \label{maping}
\end{eqnarray}
This leads to a semi-explicit scheme,
\begin{equation}
g_{t + \delta t} = g_t  + \frac{{2 \delta t }}{{\delta t + 2\tau
}}[f_t^{\eq} - g_t] + \frac{{2\tau \delta t }}{{\delta t +
2\tau}}[\Psi_t(f)  + \Phi_t(f) ]. \label{discreq}
\end{equation}
Note that the simple scheme (\ref{discreq}) is semi-explicit due
to the presence of the correction terms $\Psi_i$ and $\Phi_i$ (and
not fully explicit, as in the standard case without any
corrections). The scheme utilizes the transformed populations $g$.
The equilibrium $f^{\eq}$ can be computed using equations that
relate the locally conserved moments of the populations $f$ with
those of the $g$-populations. In order to do this, we evaluate the
moments of the transformation (\ref{maping}):
\begin{eqnarray}
\rho (f) &=& \rho (g), \label{rhoalg} \\
j_x (f) &=& j_x (g) + \frac{\delta t}{2}\sum_{i=0}^{8}c_{ix} \Psi_i (f) , \label{jxalg} \\
j_y (f) &=& j_y (g) + \frac{\delta t}{2}\sum_{i=0}^{8}c_{iy} \Psi_i (f) , \label{jyalg} \\
T(f) &=& \frac{1}{2\rho}\left( {\sum_{i=0}^{8}{c_i^2 g_i }  -
\frac{j^2(f)}{\rho} + \frac{\delta t}{2}\sum_{i=0}^{8} {c_i^2 \Phi_i (f)} }
\right) \label{Talg}
\end{eqnarray}

The above set of equations can be simplified using Eqs.\
(\ref{PsiResult}) and (\ref{PhiResult}):
\begin{eqnarray}
\rho (f) &=& \rho (g), \label{rhoalg2} \\
j_x (f) &=& j_x (g) + \frac{\delta t}{2}\left( \partial_\gamma P_{x\gamma }^{\errneq}(f) \right) , \label{jxalg2} \\
j_y (f) &=& j_y (g) + \frac{\delta t}{2}\left( \partial_\gamma P_{y\gamma }^{\errneq}(f) \right) , \label{jyalg2} \\
T(f) &=& \frac{1}{2\rho}\left( {\sum_{i=0}^{8}{c_i^2 g_i }  -
\frac{j^2(f)}{\rho} + \frac{\delta t}{2} \partial _\alpha(q_\alpha^{\err}(f)+q_\alpha ^{\errneq}(f)) } \right) . \label{Talg2}
\end{eqnarray}
Finally, discretization in space is done as in the standard
lattice Boltzmann method: if $x$ is a grid node then also $x\pm
c_i\delta t$ is the grid node.

In the simulation, the following algorithm was implemented for the
collision step:
\begin{enumerate}
\item[Step 1.] Calculate $\rho$, $j_\alpha$, $T$ using
(\ref{rhoalg2}-\ref{Talg2}), $\left( \partial_\gamma P_{\alpha \gamma }^{\errneq} \right)^{t-1}$, $\left( \partial _\alpha(q_\alpha^{\err}+q_\alpha ^{\errneq})\right)^{t-1}$  and
$g_t$ values;
 \item[Step 2.] Calculate $\left( \partial_\gamma P_{\alpha \gamma }^{\errneq} \right)^{t}$, $\left( \partial _\alpha(q_\alpha^{\err}+q_\alpha ^{\errneq}) \right) ^{t}$ using
values of $\rho$, $j_\alpha$, $T$ from Step 1;
 \item[Step 3.] Calculate
again $\rho$, $j_\alpha$, $T$ using (\ref{rhoalg2}-\ref{Talg2}),
$g_t$ and the values calculated in Step 2;
 \item[Step 4.] Use $\rho$, $j_\alpha$, $T$ from Step 3 for
the calculation of the equilibrium values  (\ref{eqpop});
 \item[Step 5.]
Use $\left( \partial_\gamma P_{\alpha \gamma }^{\errneq} \right)^{t}$, $\left( \partial _\alpha(q_\alpha^{\err}+q_\alpha ^{\errneq}) \right) ^{t}$ from Step 2 in the discrete
equation (\ref{discreq}) along with the equilibrium values
calculated in Step 4.
\end{enumerate}
The terms $\partial_\gamma P_{\alpha \gamma }^{\errneq}$ and
$\partial _\alpha(q_\alpha^{\err}+q_\alpha ^{\errneq})$ are
evaluated using a second-order accurate finite-difference scheme.

A few comments on the computational efficiency of our model are in
order. First, the present thermal model is only about $2.7$ times
slower as compared to the standard isothermal LB method on the
same $D2Q9$ lattice due to
evaluations of the deviation terms (\ref{pygamma}) and
(\ref{qaerr}).
Note that, depending on the desired accuracy and the optimization
level, faster and simpler algorithms can be readily designed. For
example, for the natural convection problem considered below in
section \ref{SecNum}, deviations of the order $j^3$, $j^2\Delta
T$, and $j^4$ can be safely neglected in the energy equation,
which removes the computational overhead with respect to the
standard isothermal $D2Q9$ code without sacrificing the accuracy
of the results in the thermal case. It should be stressed that
optimization of the code was not pursuit in the present study so
that the above estimates are only qualitative.

Second, as compared to the thermal model on two lattices
\cite{HeChenDoolen98} in terms of memory use, the present model
requires storage of twelve values per node, that is,  nine for the
populations $g_i$, two for the terms $\left(
\partial_\gamma P_{\alpha \gamma }^{\errneq} \right)^{t}$, and one
for the term $\left(
\partial _\alpha(q_\alpha^{\err}+q_\alpha ^{\errneq}) \right)
^{t}$. In the case of the two distribution function thermal models
\cite{HeChenDoolen98}, the storage of at least eighteen values per
node is required, without taking into account the additional
computational effort needed in order to recover viscous
dissipation terms.

These estimates show that the present one-lattice thermal model is
viable. We shall now proceed with a numerical validation of the
present scheme.

\section{Numerical examples}\label{SecNum}
In this section,  the thermal model developed above is validated
numerically. Equilibrium populations are expanded till the fourth
order in the velocity and the corrections described in previous
sections are applied. The same model is used in all simulations,
even though viscous heat dissipation could have been ignored in
the Rayleigh-B\'{e}nard setup.
\subsection{Necessity of guided equilibrium}
In our first example, we compare the three different models: the
original consistent LB model \cite{Ansumali05,Prasianakis06}, the
model using the guided equilibrium (\ref{eqpop}) without any
correction terms, and the full model after the correction terms
are applied. In order to demonstrate the necessity of the guided
equilibrium, we simulate the two-dimensional Couette flow between
two parallel moving plates at different temperatures.
 Isothermal walls are separated by a distance $H$ and move parallel to the $x$-axis.
 The diffusive wall boundary conditions \cite{Ansumali02} is implemented for
the isothermal walls, and periodic boundary condition is applied
for the rest. Analysis similar to the one presented in
\cite{CouettePRL07} results in the following relation at the
steady state:
\begin{eqnarray}
    P-N^{\rm eq}={\rm const},  \label{pmneqc}
\end{eqnarray}
where $P\equiv P^{\rm eq}=P_{xx}+P_{yy}$ is the trace of the
pressure tensor and
 $N=P_{xx}-P_{yy}$ is the normal stress difference.
 On one hand, using the guided equilibrium (\ref{eqpop}) in
(\ref{pmneqc}) results in the constant pressure in the whole
domain:
\begin{eqnarray}
\rho T = {\rm const}. \label{solBoltz}
\end{eqnarray}
This is consistent with the correct Maxwell-Boltzmann relation for
the pressure tensor.
 On the other hand, as we have already mentioned in section \ref{SecLE},
  the equilibrium normal stress difference
$N^{\rm eq}$ of the consistent LB model
\cite{Ansumali05,Prasianakis06} exhibits a deviation of the order
$\sim j^2 \Delta T$, where  $\Delta T$ is a variation around the
reference temperature $T_0=1/3$. In this case, equation
(\ref{pmneqc}) yields a different result, namely,  the pressure
$p=\rho T$  varies along the y-axis according to the variation of
the momentum,
\begin{eqnarray}
\rho T - \frac{{(1 - 3T)}}{{4T}}\frac{{j_x^2 }}{\rho }  = {\rm
const}. \label{solELBM}
\end{eqnarray}
\begin{figure}[htbp]
    \centering
        \includegraphics[width=0.50\textwidth]{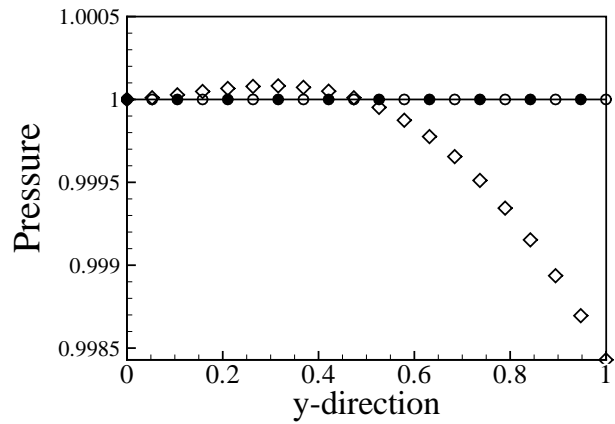}
    \caption{Pressure $p=\rho T$ as a function of the reduced $y$
    coordinate ($y_{\rm norm}=y/H$, $H$ is the distance between the plates) in the thermal Couette flow. Line: analytical
    solution (\ref{solBoltz}); Open circles: LB model with guided
    equilibrium; Filled circles: LB model with guided equilibrium
    and correction terms. Diamonds: LB model
    \cite{Ansumali05,Prasianakis06}. Reynolds number ${\rm Re}=320$. }
    \label{Fig2}
\end{figure}
In Fig.\ \ref{Fig2}, simulation results for the pressure in the
Couette flow with all the three models are presented and compared
with the analytical solution (\ref{solBoltz}). While the variation
of the pressure $p=\rho T$ away from the constant value is
numerically small for the original  model
\cite{Ansumali05,Prasianakis06}, it is still visible, and it
verifies the analytical solution (\ref{solELBM}) rather than
(\ref{solBoltz}). On the other hand, the LB model with the guided
equilibrium verifies the constancy of the pressure with machine
precision. Same holds also for the LB model with the guided
equilibrium when all the correction terms are applied, as
expected. This demonstrates necessity of the guided equilibrium in
our construction.

\subsection{Viscous heat dissipation}
\begin{figure}[tbp]
    \centering
        \includegraphics[width=0.45\textwidth]{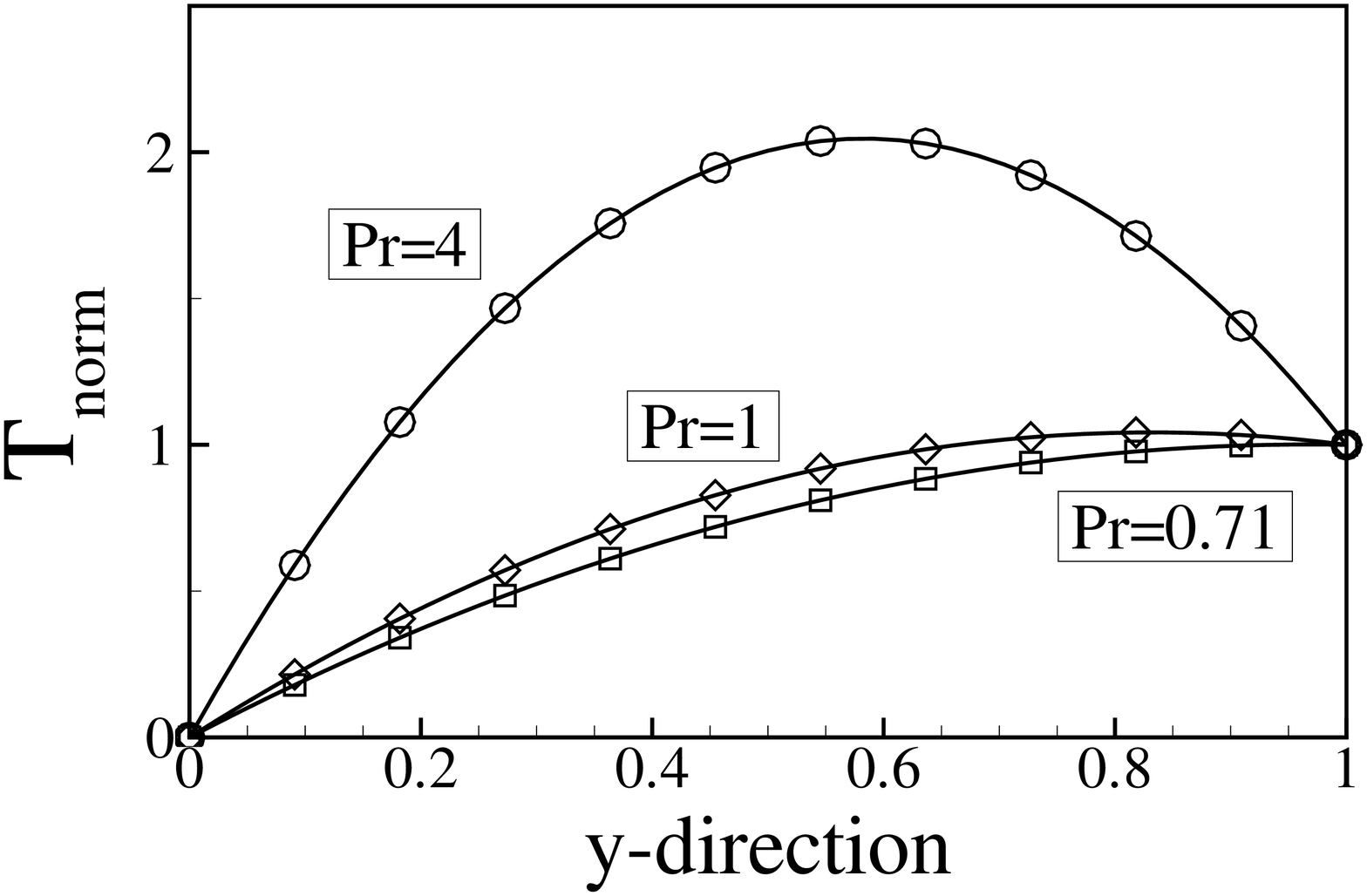}
    \caption{Temperature profile in the thermal Couette flow.
    Plotted is the reduced temperature, $T_{\rm norm}=(T-T_{\rm cold})/(T_{\rm hot}-T_{\rm
    cold})$, versus the reduced $y$-coordinate (as in Fig.\
    \ref{Fig2}). Eckert number $\Ec=3$.
    Line: Analytical solution for $\Pr=4$, $\Pr=1$, and $\Pr=0.71$. Reynolds number ${\rm Re}=200$.
    Symbol: Simulation results for the three cases (see text).  Diamonds: Case
    1; Squares: Case 2 (air); Circles: Case 3.
    }
    \label{Fig3}
\end{figure}

We simulate the thermal Couette flow as in the preceding section
but for a small temperature differences. In this case, the viscous
heat dissipation is important and affects the temperature profile.
As is well known, the Navier-Stokes and Fourier equations
(\ref{NSF}) predict that the temperature profile depends on the
Prandtl number, $\Pr$, and on the Eckert number, $\Ec$, where
\begin{equation}
\Ec=\frac{u^2}{c_p \Delta T}, \label{Ec}
\end{equation}
with $u$  the difference of the velocities of the plates, and
$\Delta T$  the difference of temperatures of the plates.
This simulation enables to verify that the correction terms do recover
the right Prandtl number. In the simulation, we used $\Ec=3$ and
$\Delta T=2 \text{x} 10^{-4}T_{0}$, where  $T_0=1/3$ is the reference temperature.
Three different cases were considered:

\begin{enumerate}
\item[Case 1.] LB with the guided equilibrium (\ref{eqpop}) and
the corrections $\Psi_i$ and $\Phi_i$ which deliver $\Pr=1$ (that
is, without the transformation (\ref{prcorr});
\item[Case 2.]  As
in Case 1 plus the transformation (\ref{prcorr}) to match
$\Pr=0.71$ (air).
\item[Case 2.]As in Case 1 plus the
transformation (\ref{prcorr}) to match $\Pr=4$.

\end{enumerate}
Simulation results for the temperature profile are presented in
Fig.\ \ref{Fig3}, and are in excellent agreement with the
analytical solution. For the current setup, with the present
implementation of boundary condition, simulation results remain in
agreement with the analytical solution for Prandtl numbers ranging
from $0.01$ to $20$.

\subsection{Rayleigh-B\'{e}nard convection}
The Rayleigh-B\'{e}nard convection flow is a classical benchmark
on the thermal models. The fluid is enclosed between two parallel
stationary walls, the hot (bottom) and the cold (top), and
experiences the gravity force. Density variations caused by the
temperature variations drive the flow, while the viscosity will
counteract to equilibrate it.

In our LB model, gravity is implemented using the correction
(\ref{gravcorr}). We operate the model in the weakly compressible
regime, however, without using the Boussinesq approximation. For
that we set a temperature difference of the order of $3\%$ of the
reference temperature $T_0=1/3$. At the top and the bottom walls
we apply the diffusive wall boundary condition \cite{Ansumali02},
whereas periodic boundary condition is applied on the vertical
walls. For the nodes that belong to the isothermal walls,
gravitational force was not applied.

The Prandtl number used corresponds to air, $\Pr=0.71$. The
dimensionless number that characterizes the flow is the Rayleigh
number $\Ra$,
  \begin{equation}
  \Ra = \frac{\Pr g\beta \Delta TH^3}{\nu ^2},
  \label{Ra}
  \end{equation}
   where $g$ is the gravity acceleration,
   $\Delta T$ is the wall temperature difference,
   $H$ is the distance between the walls and
   $\nu$ is the kinematic viscosity of the fluid.
For the ideal gas equation of state, the thermal expansion
coefficient $\beta$ is defined as:
\begin{eqnarray}
\beta  =  - \frac{1}{\rho }\left( {\frac{{\partial \rho
}}{{\partial T}}} \right)_P = \frac{1}{T}.
\end{eqnarray}

For the computation of the Rayleigh number, we used the thermal
expansion coefficient evaluated at the reference temperature
$T_0=1/3$, that is, $\beta=3$. The heat transfer is described by
the Nusselt number, $\Nu$, defined as the ratio between convective
heat transport to the heat transport due to temperature
conduction:
\begin{eqnarray}
    \Nu = 1 + \frac{{\left\langle {u_y T} \right\rangle H }}{{\kappa \Delta T}}.
\end{eqnarray}
Here $\left\langle {u_y T} \right\rangle$ denotes the average over
the convection layer and $\kappa$ is the thermal conductivity of
the model (\ref{transportcoefficients}).  For the current
simulations, a computation domain with the aspect ratio $2:1$ was
considered.
\begin{figure}
\centering
\epsfig{file=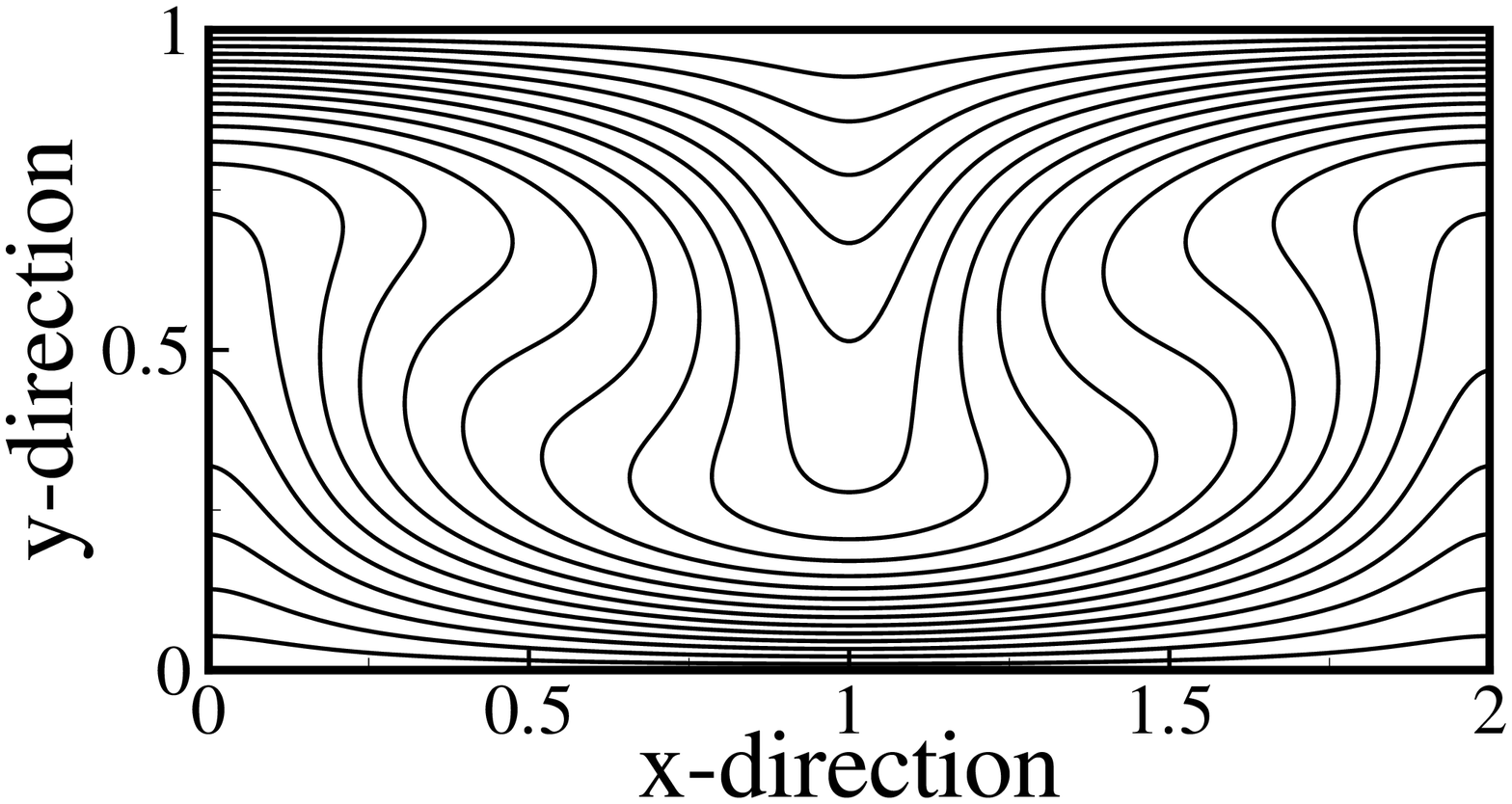,width=0.7\linewidth,clip=}\\
\epsfig{file=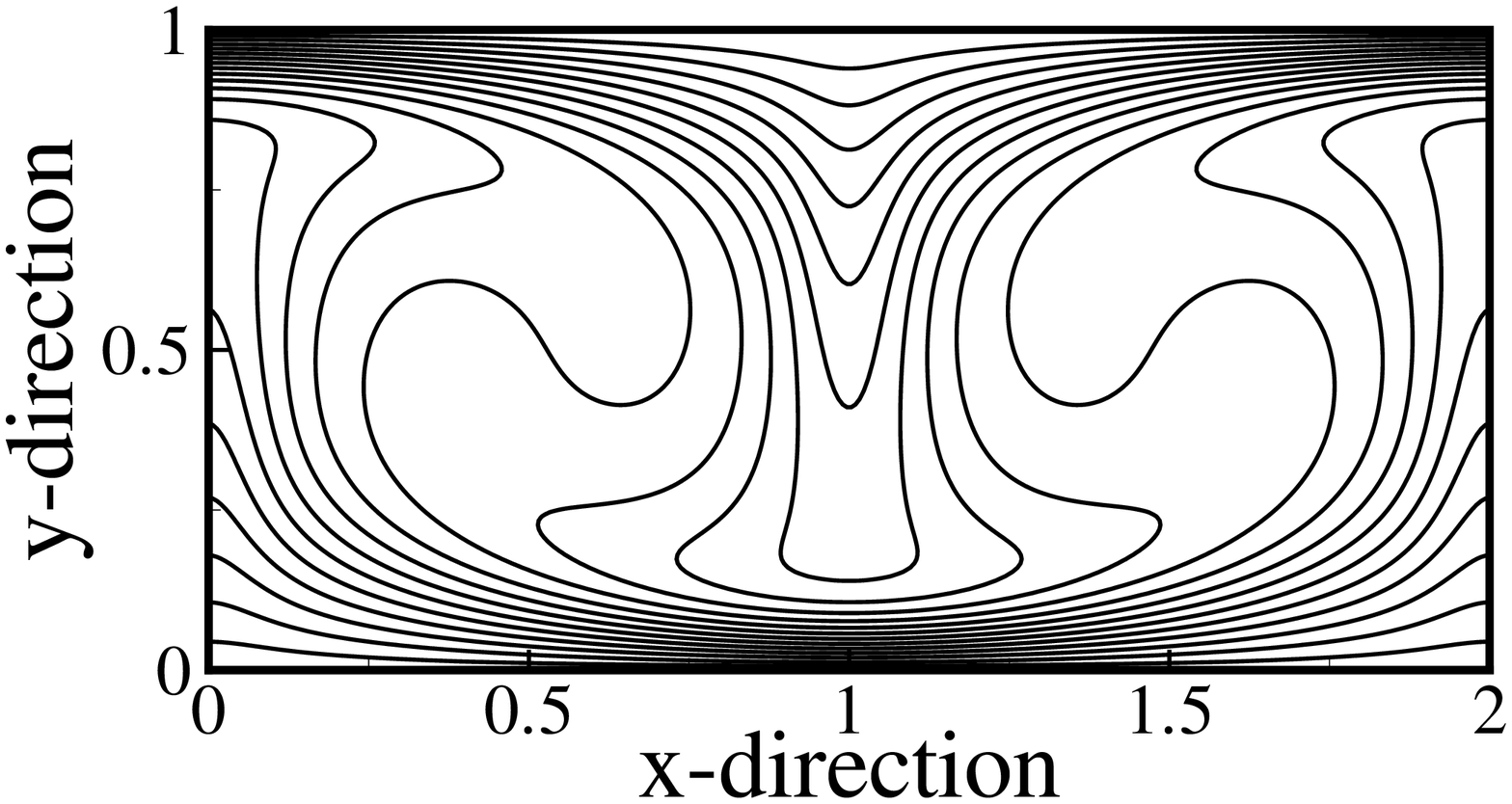,width=0.7\linewidth,clip=}
\caption{Contour plot with $19$ equidistant iso-temperature lines
for $\Ra=10^4$ (top) and $\Ra=5\text{x}10^4$ (bottom).}
\label{Fig4}
\end{figure}

\begin{figure}
\centering
\epsfig{file=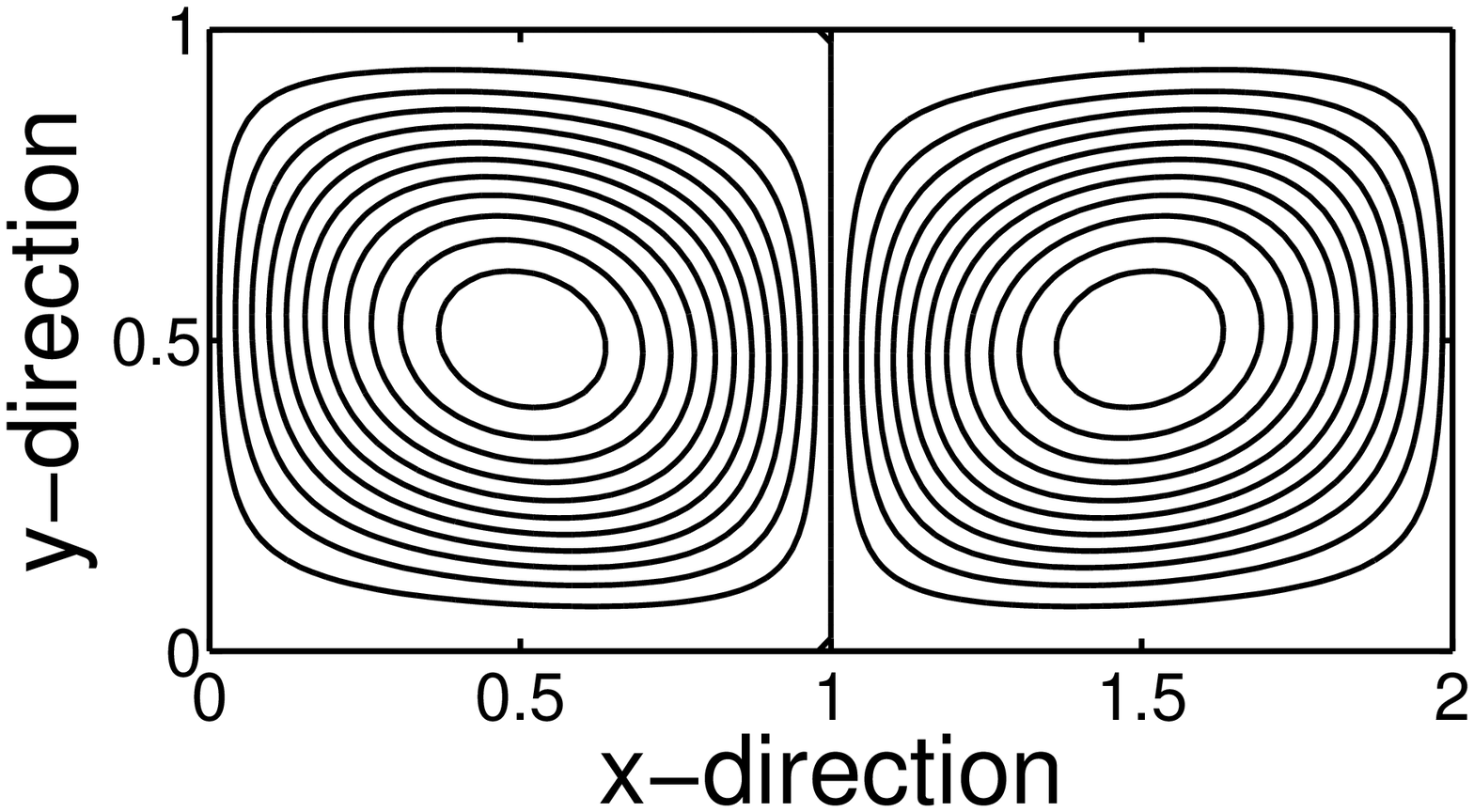,width=0.7\linewidth,clip=}\\
\epsfig{file=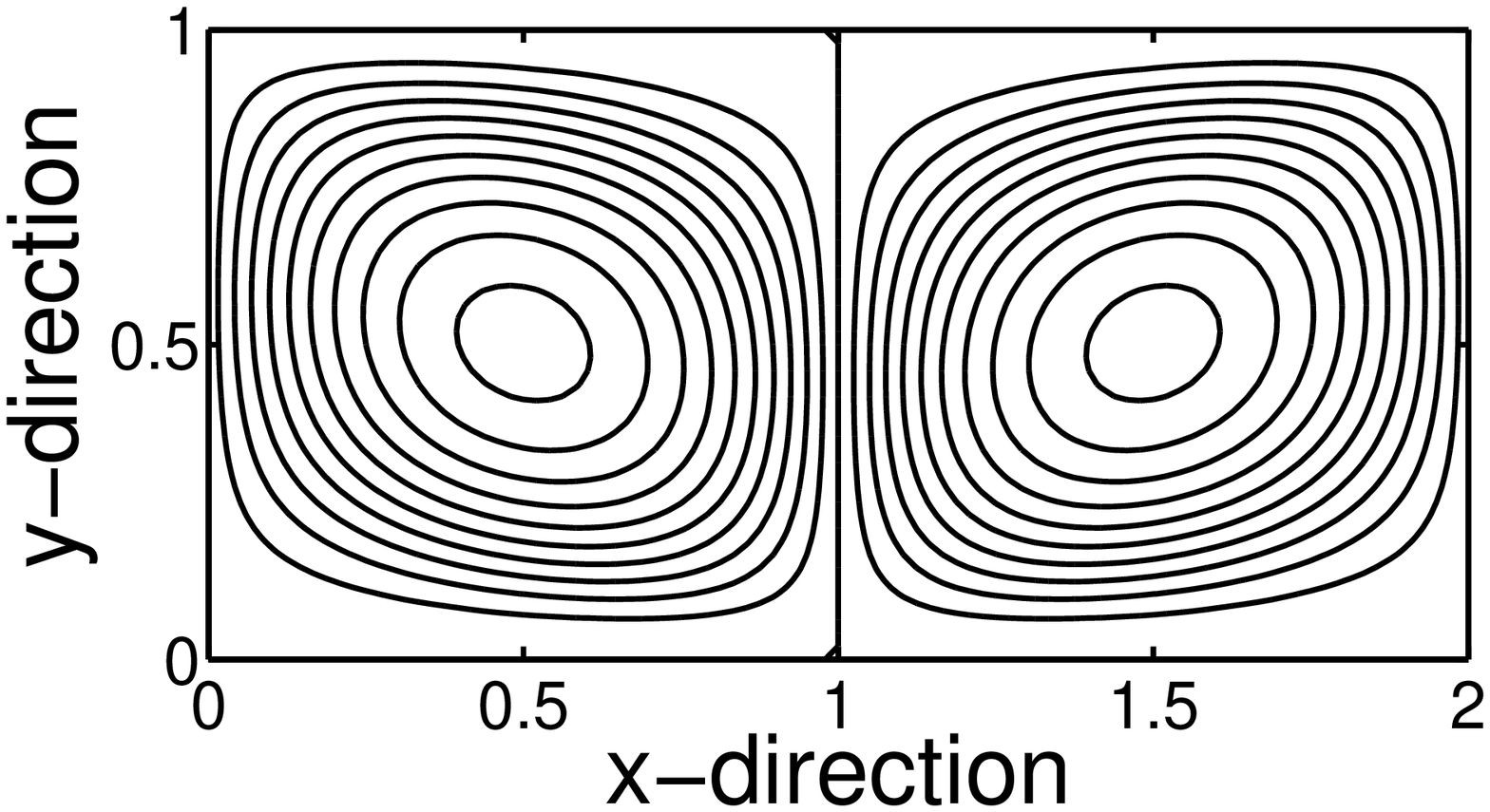,width=0.7\linewidth,clip=}
\caption{Stream function contours for $\Ra=10^4$ (top) and
$\Ra=5\text{x}10^4$ (bottom). Difference between contour lines is
$0.025$ in units of $\rho_{max}U_{max}H$.} \label{Fig5}
\end{figure}
For this setup, the critical Rayleigh number was found to be
$\Ra_{\rm cr}=1700 \pm 10$, which is consistent with the reference
data for the Boussinesq approximation
\cite{CleverBusse74,Busse78}. Any flow field was dissipated for
values less than $\Ra_{\rm cr}$. On the contrary, for values
larger than $\Ra_{\rm cr}$, the flow develops two or more cells
(vortices) depending on the initial conditions.

\begin{figure}[tbp]
\centering
\includegraphics[width=0.450\textwidth]{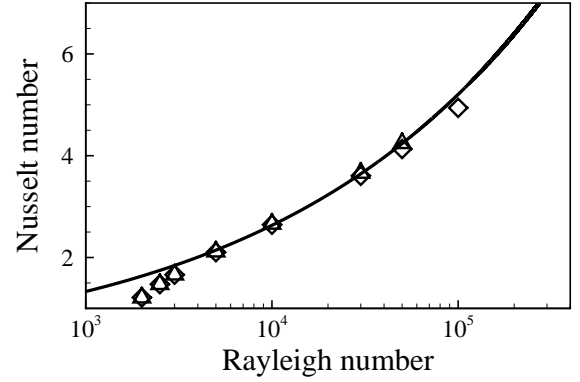}
\caption{Rayleigh B\'{e}nard convective flow. Nusselt number vs.
Rayleigh number. Diamonds: The current LB model; Triangles:
Reference data of Ref.\ \cite{CleverBusse74}; Line: Empirical
power-law $\Nu=1.56(\Ra/\Ra_c)^{0.296}$ \cite{CleverBusse74}.}
\label{Fig6}
\end{figure}

\begin{figure}
\centering
  \includegraphics[width=0.450\textwidth]{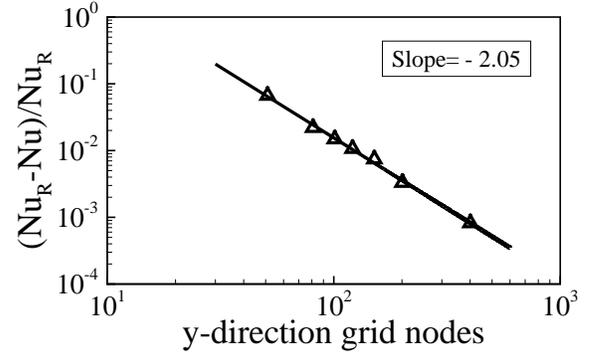}
\caption{Grid convergence study for $\Ra=3\text{x}10^4$. Symbol: The
relative error between the simulation result $\Nu$ and the
converged solution $\Nu_{\rm R}$. Line is a fitted curve which
reveals second order convergence.} \label{Fig7}
\end{figure}

For different Rayleigh numbers, simulations for the $51$, $101$
and $201$ grid nodes in the $y$-direction were performed.
Extrapolating the obtained values of Nusselt number, an estimate
of the final converged solution can be done ($\Nu_{\rm R}$). In
Fig.\ \ref{Fig4}, the isotherms of $\Ra=10^4$
 and $\Ra=5\text{x}10^4$ are plotted for the case of $101$ grid nodes in the y-direction.
 In Fig.\ \ref{Fig5}, the contours of the stream function of the compressible flow field
  for $\Ra=10^4$ and $\Ra=5\text{x}10^4$ are plotted.
The extrapolated converged values of Nusselt number at various
Rayleigh numbers are plotted in Fig.\ \ref{Fig6}, and compared
with the standard reference data \cite{CleverBusse74,Busse78} in
Table \ref{table1}.
 The present model is found to be in good agreement with \cite{CleverBusse74}, and it
also agrees well with the thermal LB models on double lattices
where the temperature dynamics is treated as a passive scalar
\cite{HeChenDoolen98,Shan97}. It should be noted that for $\Ra$ up
to $10^4$ a uniform grid with $51$ nodes in the $y$-direction was
sufficient, while for larger values of $\Ra$, larger grids provide more accurate results.
 In Fig.\ \ref{Fig7}, a grid convergence study is performed.
Natural convection of $\Ra=3\text{x}10^4$ in the same setup is considered.
 Fig.\ \ref{Fig7} reveals the second-order accuracy of the numerical
scheme.

\begin{widetext}
\begin{table}
\centering
\begin{tabular}{|c|c|c|}
\hline
 $\text{ Rayleigh number }$ & $\text{ Nusselt number ($\Nu_{\rm R}$) }$ & $\text{Difference
 from Ref. \cite{CleverBusse74}}$ \\
\hline
2500&1.474&0.07 \% \\ \hline
5000&2.104&0.57 \% \\ \hline
10000&2.644&0.64 \% \\ \hline
30000&3.605&1.56 \% \\ \hline
50000&4.133&2.64 \% \\
\hline
\end{tabular}
\caption{Simulation results of the present  model compared to the
simulation results of Ref.\ \cite{CleverBusse74}. Left column:
Rayleigh number; Central column:  Nusselt number; Right column:
Difference in percent from the values of Ref.\
\cite{CleverBusse74}.} \label{table1}
\end{table}
\end{widetext}

\begin{figure}
\centering
  \includegraphics[width=0.50\textwidth]{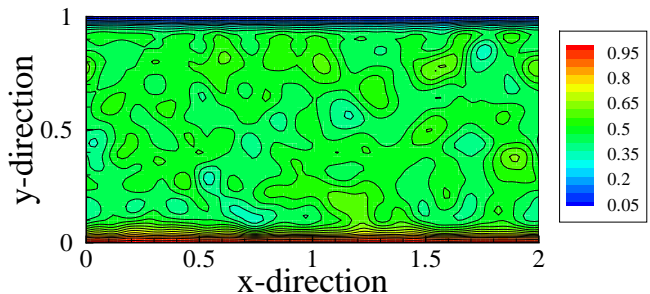}
\caption{(Color online) Snapshot of the temperature field at
$\Ra=10^8$, $\Pr=0.71$, using a uniform grid of 51 nodes for the
y-direction. Contours of reduced temperature, $T_{\rm
norm}=(T-T_{\rm cold})/(T_{\rm hot}-T_{\rm cold})$ are plotted.}
\label{Fig8}
\end{figure}

Finally, it should be mentioned that on a grid with only $51$
nodes in the $y$-direction, the code was run stably for $\Ra$ as
high as at least $10^8$ which proves a good numerical stability of
the algorithm. In Fig.\ \ref{Fig8}, a snapshot of the temperature
field is presented at $\Ra=10^8$. However, a study of the natural
convection at high Rayleigh numbers is out of the scope of this
paper.

\section{Summary}\label{SecSum}
A reader who wishes to implement this model should execute  the
following steps:
\begin{itemize}
    \item Use guided equilibrium populations (\ref{eqpop});
    \item Calculate the correction terms $\Psi_i$ and $\Phi_i$
    with a second-order finite-difference scheme.
    \item Use the discretized equation (\ref{discreq}).
\end{itemize}

\section{Conclusion}\label{SecConcl}

In this paper, we have developed a new  lattice Boltzmann model
for simulation of thermal flows with the standard and most
commonly used $D2Q9$ lattice. The important starting point of the
derivation is the consistent LB method \cite{Ansumali05}. Unlike
the previous approaches, the consistent LB model on standard
lattices already includes energy as a locally conserved quantity,
the nonequilibrium stress tensor is free of a spurious bulk
viscosity, and a part of the moment relations satisfies the
required Maxwell-Boltzmann relations. Hence, by using the
consistent LB model, it becomes unnecessary to introduce a
separate lattice for the energy field.

We have considerably refined the consistent LB model in two steps.
First, following the concept of guided equilibrium
\cite{Karlin98}, we recovered Galilean invariance of the pressure
tensor at non-vanishing variations of the temperature. Second, we
have identified the remaining deviations in the
Navier-Stokes-Fourier equations, and have removed them by adding
compensating terms into the kinetic equation. This step allowed us
also to make the Prandtl number a tunable parameter, and to add
external forces. It should be noted that additional terms also
appear in the method based on two lattices \cite{HeChenDoolen98}
but there is no relation between the both. On the other hand, a
numerical implementation of the kinetic equation with additional
terms here has practically not much different from the method
first developed in \cite{HeChenDoolen98}. Finally, it is
straightforward to extend the present model to three dimensions
based on the $D3Q27$ consistent LB model \cite{Ansumali05}.

We shall conclude this paper with a general comment on the
construction of thermal lattice Boltzmann models. As we already
mentioned in the introduction, there are two basic directions for
such a development, one which uses larger and more isotropic sets
of velocities, and the other which uses a small number of
velocities and introduces non-local (dependent on the gradients of
the fields) corrections to compensate for the deficit of lattice
symmetry. This situation resembles the classical dilemma of
extending the hydrodynamics into beyond the Navier-Stokes level:
One way is the higher-order hydrodynamics (Burnett or
super-Burnett) which means higher-order derivatives and more
non-locality in the equations for  the standard hydrodynamic
fields. The other is to take more moments while staying local
(Grad's moment method). Both approaches have positive and negative
aspects. In our case, the attractive features of large velocity
sets (locality, isotropy, possibility of the exact treatment of
propagation) should be waged against the necessity of handling
many more fields (populations). On the other hand, the obvious
attractive side of the small lattices approach has to be opposed
by the fact that one has to evaluate more terms dependent on the
gradients of the fields. While this proves to be possible, still,
one needs to be careful about numerical errors brought in by such
procedures. We believe that both these approaches, at present,
show there advantages and disadvantages, so the decision about
which to prefer should be the focus of future studies.

We are indebted to our colleagues S. Ansumali, S. Arcidiacono, K.
Boulouchos, S. Chikatamarla, C. Frouzakis and J. Mantzaras for
their help, questions and discussions. We gratefully acknowledge
support by BFE Project No. 100862, CCEM-CH and by ETH Project No.
0-20235-05.

\appendix
\section{Derivation of hydrodynamic equations}
\label{AppendixA}

We apply the Chapman-Enskog expansion in order to derive
hydrodynamic equations corresponding to the BGK equation
(\ref{KE1}) with the equilibrium (\ref{eqpop}) Populations
 $f_i$ and the time derivative operator are expanded into powers of
 the relaxation parameter $\tau$
\begin{eqnarray}
f_i  &=& f_i^{eq}  + \sum\nolimits_{n = 1}^\infty  f_i ^{(n)} \tau ^n, \\
\partial _t &=& \sum\nolimits_{n = 0}^\infty  {\tau ^n \partial _t^{(n)}
}.
\end{eqnarray}
The non-equilibrium pressure tensor and energy flux to first order
in $\tau$ are defined as:
\begin{equation}
P_{\alpha \beta }^{(1)}  = \tau \sum\limits_{i = 0}^8 {c_{i\alpha
} c_{i\beta } f_i^{(1)} },\ q_\alpha ^{(1)}  = \tau \sum\limits_{i
= 0}^8 {c_{i\alpha } c_i^2 f_i^{(1)} }.
\end{equation}
On the zeroth order, equations for density, momentum, pressure
tensor and energy flux are:
\begin{eqnarray}
\partial_t^{(0)} \rho &=& - \partial_{\alpha}\, j_{\alpha}, \label{dtrho}    \\
\partial _t^{(0)} j_\alpha  &=& - \partial _\beta   {P_{\alpha \beta }^{\MB}}, \label{dtja} \\
\partial _t^{(0)} \left[ {\frac{{j^2 }}{\rho } + 2\rho T} \right] &=&  -
\partial _\alpha q_\alpha^{\MB}- \partial _\alpha q_\alpha^{\err}, \label{CE2} \\
\partial _t^{(0)} P_{\alpha \beta }^{\eq} &=& - \partial _\gamma  Q_{\alpha \beta \gamma }^{\eq}
-\frac {1} {\tau} P_{\alpha \beta }^{(1)},\label{MainPressure}\\
\partial _t^{(0)} q_{\alpha}^{\eq} &=& - \partial _\beta  R_{\alpha \beta}^{\eq}
-\frac {1} {\tau} q_{\alpha}^{(1)},\label{MainQ}
\end{eqnarray}
where $Q_{\alpha \beta \gamma }^{\eq}$ and $R_{\alpha
\beta}^{\eq}$ are the third- and the fourth-order moments
evaluated at the local equilibrium (\ref{eqpop}) (explicit form of
these function was given in section \ref{SecLE}). From
(\ref{CE2}), we derive the zeroth-order equation for the
temperature,
\begin{eqnarray}
 \partial _t^{(0)} T &=& -T\partial _\alpha \left(\frac{{j_\alpha }}{\rho }\right)-
 \frac{{j_\alpha }}{\rho }\partial _\alpha T - \frac{1}{{2 \rho}}\partial _\alpha
 q_\alpha^{\err}.
\end{eqnarray}
On the  first order in $\tau$ we need  equations for the locally
conserved fields only,
\begin{eqnarray}
\partial_t^{(1)} \rho &=& 0,    \label{1orderDensity}                          \\
\partial_t^{(1)} j_{\alpha}  &=& - \frac {1} {\tau} \partial_{\beta}\, P_{\alpha\beta}^{(1)}, \label{1orderMomentum} \\
\partial _t^{(1)} \left[ {\frac{{j^2 }}{\rho } + 2\rho T} \right] &=&  - \frac {1} {\tau} \partial _\alpha q_\alpha^{(1)}. \label{CE3}
\end{eqnarray}
Function $P_{\alpha \beta }^{(1)}$ is represented as a sum of two
terms, $P_{\alpha \beta }^{\neqq}$, which results in the correct
Navier-Stokes term in the momentum equation, and $P_{\alpha \beta
}^{\errneq}$, representing the deviation:
\begin{eqnarray}
P_{\alpha \beta }^{(1)}&=& P_{\alpha \beta }^{\neqq}+P_{\alpha \beta }^{\errneq}, \\
P_{\alpha \beta }^{(1)} &=&  - \tau \left[ \partial _\gamma  Q_{\alpha \beta \gamma }^{\eq}
 + \partial _t^{(0)} P_{\alpha \beta }^{\eq} \right],\\
P_{\alpha \beta }^{\neqq} &=&  - \tau \left[ \partial _\gamma
Q_{\alpha \beta \gamma }^{\MB}  + \partial _t^{(0)} P_{\alpha
\beta }^{\MB} \right].
\end{eqnarray}
In order to derive $P_{\alpha \beta }^{(1)}$ from
(\ref{MainPressure}), we note that the left hand side can be
computed by chain rule,
\begin{eqnarray}
\partial_t^{(0)} P_{\alpha \beta }^{\eq}  &=&\frac{{\partial P_{\alpha \beta }^{\eq} }}{{\partial \rho }}\partial _t^{(0)} \rho
+ \frac{{\partial P_{\alpha \beta }^{eq} }}{{\partial
j_{\gamma}}}\partial _t^{(0)} j_{\gamma}\nonumber\\&& +
\frac{{\partial P_{\alpha \beta }^{\eq} }}{{\partial T}}\partial
_t^{(0)} T. \label{dtpab}
\end{eqnarray}
Using (\ref{dtrho}), (\ref{dtja}), (\ref{CE2}) and (\ref{dtpab})
in (\ref{MainPressure}), we obtain
\begin{eqnarray}
P_{\alpha \beta }^{\neqq}&=&- \tau \rho T\!\left[\partial _\alpha
\left(\frac{{j_\beta  }}{\rho }\right)\! +\! \partial _\beta
\left(\frac{{j_\alpha }}{\rho }\right)\!-\!
\partial _\gamma \left(\frac{{j_\gamma }}{\rho }\right)\delta
_{\alpha \beta } \right],
\label{RightPressure}\\
P_{\alpha \beta }^{\errneq} &=&
\tau\left(\frac{1}{2\rho}\right)\frac{{\partial P_{\alpha \beta
}^{\eq} }}{{\partial T}}\partial _\gamma q_\gamma^{\err} -\tau
\partial _\gamma  Q_{\alpha \beta \gamma
}^{\err}.\label{PressureError}
\end{eqnarray}
Substituting (\ref{RightPressure}) and (\ref{PressureError}) into
(\ref{1orderMomentum}), and combining the latter with
(\ref{dtja}), and also combining (\ref{dtrho}) with
(\ref{1orderDensity}), we derive the equations for the density and
velocity $u_{\alpha}=j_{\alpha}/\rho$ to first order,
\begin{eqnarray}
\partial _t \rho  &=&  - \partial_{\alpha}(\rho u_{\alpha}), \\
\partial _t u_\alpha  &=&  - u_\beta\partial_\beta u_\alpha
- \frac{1}{\rho }\partial_{\alpha}( \rho T) - \frac{1}{\rho
}\partial_\beta\Pi _{\alpha \beta}\nonumber\\&& - \frac{1}{\rho
}\partial _\gamma  P_{\alpha \gamma }^{\errneq}, \label{MomApp}
\end{eqnarray}
where
\begin{equation}
 \Pi _{\alpha \beta}  = -\tau \rho
T\left[\partial_\beta u_\alpha + \partial_\alpha u_\beta -
(\partial_\gamma u_\gamma)\delta _{\alpha \beta} \right],
\end{equation}
is the nonequilibrium pressure tensor of a Newtonian fluid, and
$P_{\alpha \gamma }^{\errneq}$ (\ref{PressureError}) is the
deviation. Expanding Eq.\  (\ref{PressureError}), we arrive at the
explicit form of the deviation, Eq.\ (\ref{pygamma}).

Derivation of the equation for the energy (or, equivalently, for
the temperature) is done in a similar way upon computing the
first-order correction $q_{\alpha}^{(1)}$ from equation
(\ref{MainQ}). The resulting equation for the temperature reads:

\begin{eqnarray}
\partial _t T &=&  - u_\alpha \partial _\alpha T - T\partial _\alpha u_\alpha
\nonumber\\
&& - \frac{1}{\rho }\left[ {\partial _\alpha (2\tau \rho T\partial
_\alpha T) - \left( \partial _\beta  u_\alpha \right) \Pi _{\alpha
\beta } } \right]\nonumber\\
&& - \frac{1}{2\rho }\partial _\kappa  q_\kappa ^{\err}
-\frac{1}{2\rho } \partial _\gamma q_{\gamma}^{\errneq}.
\label{TemperatureApp}
\end{eqnarray}
The last two terms in this equation represent the deviation. Note
that, unlike in the case of the momentum equation (\ref{MomApp}),
the deviation also includes a zero-order  term $q_\kappa ^{\err}$.
This happens because, with the choice of the guided equilibrium
(\ref{eqpop}), we have set the equilibrium pressure tensor in the
required Maxwell-Boltzmann form but not the equilibrium energy
flux (why the latter is impossible on the $D2Q9$ was explained in
section \ref{SecLE}). Finally, the first-order deviation (last
term in (\ref{TemperatureApp})) has the form given by Eqs.\
(\ref{qaerr}) and (\ref{EnoneqShort}).

\bibliography{ThermalLB}

\end{document}